\documentstyle[11pt, newpasp,twoside,epsf]{article}
\def\lsim {\lower .1ex\hbox{\rlap{\raise .6ex\hbox{\hskip .3ex
        {\ifmmode{\scriptscriptstyle <}\else
                {$\scriptscriptstyle <$}\fi}}}
        \kern -.4ex{\ifmmode{\scriptscriptstyle \sim}\else
                {$\scriptscriptstyle\sim$}\fi}}}
\def\eps@scaling{.95}
\def\epsscale#1{\gdef\eps@scaling{#1}}
 \def\plotone#1{\centering \leavevmode
\epsfxsize=\eps@scaling\textwidth \epsfbox{#1}}

\begin{document}

\title{Giant planet formation from disk instability; cooling
and heating}

\author{Lucio Mayer}
\affil {Institute for Theoretical Physics, University of Z\"urich, CH-8057 Zuric
h, Switzerland}

\author{James Wadsley}
\affil {Department of Physics \& Astronomy, McMaster University, Canada}

\author{Thomas Quinn}
\affil {Department of Astronomy, University of Washington, Seattle (USA)}

\author{Joachim Stadel}
\affil {Institute for Theoretical Physics, University of Z\"urich, CH-8057 Zuric
h, Switzerland}

\begin{abstract}

We present the results of high resolution SPH simulations
of the evolution of gravitationally unstable protoplanetary disks. We
report on calculations in which the disk is evolved using a locally isothermal
or adiabatic equation of state (with shock heating), and also on new 
simulations in which cooling and heating by radiation are 
explicitly modeled. We find that disks with a minimum Toomre parameter 
$< 1.4$ fragment into several gravitationally bound protoplanets with 
masses from below to a few Jupiter masses.  This is confirmed also
in runs where the disk is given a quiet start, growing gradually in mass
over several orbital times.
A cooling time comparable to the orbital time is needed 
to achieve fragmentation, for disk masses 
in the range $0.08-0.1 M_{\odot}$.
After about 30 orbital times, merging between the bound condensations always
leads to 2-3 protoplanets on quite eccentric orbits.

\end{abstract}

The formation of gas giants as a result of fragmentation 
in a protoplanetary disk, an old idea (Kuiper 1951, Cameron 1978), has
been recently revived by a number of studies that are finally laying the
ground for a quantitative understanding of such process (Boss 2001, 2002;
Mayer et al. 2002, 2003;Pickett et al. 2000, 2003). 
The renewed consideration of this mechanism stems from several problems
faced by the conventional model of giant planet formation, in which first a 
rocky core is assembled over $10^5-10^6$ years and then 
a gaseous envelope is accreted in a few million years or more, the
exact timescale being dependent on the details of the models like 
the disk surface density and the opacity of the mixture of gas and dust
(Lissauer 1993).
These timescales are an order of magnitude too long to form planets 
before the disk is dissipated by photoevaporation in highly irradiated environments like the Orion nebula (Throop et al. 2001), and are a bit too tight 
even when compared
to the typical lifetime of disks in more quiet environments like Taurus
(Haisch, Lada \& Lada 2001). One of the strong points of the core 
accretion model, namely the
prediction that gas giants have solid cores, is considerably weakened
now that new models of the interior of Jupiter are consistent with 
the total absence of such a core (Guillot 1999).
The discovery of extrasolar planets (Marcy \& Butler 1998) 
has worsened the situation
further because now we need to explain the existence of very massive
planets, up to ten times larger than Jupiter, that in the standard model
would either take too long to be formed, might not form at all 
(Bate et al. 2003) or could migrate towards the central 
star before  being able to accrete enough mass (Nelson et al. 2000). 
The distribution of their orbital
eccentricities must also be explained; whilst in the past few years
several papers have proposed a variety of explanations, sometimes tuned
to the properties of one particular system, none of these is 
valid in general. Mayer et al. (2002, hereafter MA02) showed for 
the first time that
if a massive disk remains cold long enough, until the growing overdensities
reach some density threshold, fragmentation takes place even when heating
from compression and shocks is included ---the resulting clumps survive
for tens of orbital times, leading to systems of only a few protoplanets
with masses and orbital eccentricities in the range of observed extrasolar
planets. The strong point of the 3D smoothed particle hydrodynamics (SPH)
simulations of MA02 was the high resolution (up to
$10^6$ gas particles) that allowed to resolve the local Jeans length 
across a wide range of densities, including the regime in which
fragmentation takes place (Bate \& Burkert 1997; Nelson 2003). 
However, these simulations
where quite simplified in the thermodynamics (see Pickett et al. 2000; 2003),
since only two extreme conditions, a locally isothermal or an 
adiabatic equation of state (with shock heating), were employed.
Boss (2001, 2002), by using radiative transfer in the diffusion
approximation, has shown that efficient cooling in the disk midplane
can occur thanks to convective transport of heat to the disk
atmosphere. He finds that the timescale for convective cooling is
comparable to the orbital time in the outer, colder regions of a disk 
of mass $\sim 0.1 M_{\odot}$.
Here we review the main results obtained with the large
survey of simulations extensively
described in Mayer et al. (2003, hereafter MA03), and we present the first
results of new simulations in which radiative cooling and heating are
directly implemented.
Following recent work by Pickett et al. (2003) and
Rice et al. (2003a,b), we will investigate how fast cooling has to be
for fragmentation into gravitationally bound planets to occur.

\begin{figure}
\epsscale{0.8}
\plotone{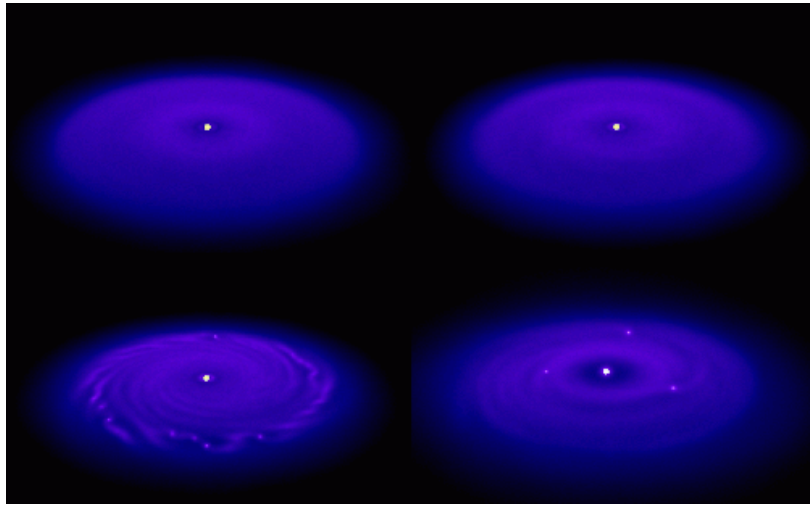}
\caption{Snapshots of the simulation in which the disk is grown in mass
(see text, section 2.1). The disk logarithmic density is shown using 
color-coding and for an inclination of 45 degrees with respect to the 
disk plane; brighter colors are used for higher densities, and
maximum densities are of order $10^{-6}$ g/cm$^3$. 
From top left to bottom right, boxes show the inner 25 AU at t=0, 
t=300 years, t=450 years and t=650 years.}
\label {fig1}
\end{figure}

\section{Initial Conditions and Simulations}

The 3D disk models initially extend between 4 and 20
AU. The central star is modeled as a softened point mass with 
a mass of $1 M_{\odot}$. There are
no fixed boundaries; the disk is free to expand and contract and both
the central star and the disk can respond to their mutual
gravitational field. The disk surface density profile is of the type
$\Sigma \sim r^{-3/2}$.
Disk models have masses between $0.075$ and $0.125 M_{\odot}$ and initial
temperature profiles as those used in MA02. The minimum
temperatures are reached at the outer edge of the disk and range from 36 to 
60 K. More details on the setup of the initial conditions are explained
in MA03. Models are
evolved with a locally isothermal or with an adiabatic equation of state;
the new runs adopt an adiabatic equation of state with the addition of
radiative cooling and heating. In all cases
we include a standard Monaghan artificial viscosity to model
shocks, with typical values of the coefficients $\alpha=1$ and 
$\beta=2$. The artificial
viscosity term appears in both the momentum and the energy equation, 
hence irreversible shock heating is included, except in the
locally isothermal equation of state, where by construction the 
thermal energy of any given particle is assumed to be constant.

Radiative cooling is implemented
using a  prescription similar to that 
used by Rice et al. (2003a) (see also Pickett et al. 2003); the 
cooling timescale is simply assumed to be proportional
to the local orbital frequency of particles, 
$t_{cool}(r)=A{\Omega(r)}^{-1}$. In addition we introduce a 
density dependent criterion, so that when a region grows beyond a
specified threshold, radiative cooling is completely switched off. In the runs
presented here the density threshold is fixed at $5 \times 10^{-10}$ g/cm$^{3}$
--- this is a conservative choice based on the recent 
calculations by Boss (2001) with radiative transfer, which show
that at such densities the temperature evolves in a nearly adiabatic
fashion. 
In runs that are evolved using a locally isothermal
equation of state we simply switch to an adiabatic equation of state
throughout the disk once such density threshold is reached somewhere
(see also MA02). 

In the runs with radiative cooling we
heat the inner part of the disk by means of another radially dependent
term (this goes exponentially to zero at $R=10$ AU) so that a
gravitationally stable disk ($Q >2$) develops a temperature profile
similar to that used in the initial conditions of the locally
isothermal runs (the latter 
was indeed motivated by the results of the radiative transfer models of Boss
(1996) that include irradiation from the central star and
compressional heating due to material infalling onto the disk
from the molecular cloud, see MA03). 

The simulations are run with GASOLINE, a parallel tree-based gravity code
with SPH hydrodynamics which uses multistepping to evolve efficiently
a wide density range (Wadsley, Stadel \& Quinn 2003).

\section{Results}

In what follows we describe the main results of our large suite of numerical
simulations, describing first the locally isothermal and purely adiabatic
runs, and then those including  radiative cooling and heating. A detailed 
description of the former can be found in MA03.

\subsection{Locally isothermal and adiabatic runs}

Disks evolved with a locally isothermal equation of state and with 
$Q_{min} < 1.4$ fragment after 6-7 orbital times (we used the orbital
time at 10 AU, 28 years, as a reference), the others ($Q_{min}=1.5-1.9$)
develop only from very strong to moderate spiral patterns which reach a 
peak amplitude after 
6-7 orbital times and then fade and saturate towards
a nearly stationary pattern (see MA02). With  $Q_{min} < 1.4$, clump
formation proceeds even when the equation of state is switched from
locally isothermal to adiabatic once the critical density threshold is
reached (see previous section), although the clumps that survive and
become gravitationally bound are fewer due to strong shock heating 
along the spiral arms (see MA03).
Clumps form on eccentric orbits (these follow the path of the spiral arms) 
at scales from below to just about
one Jupiter mass, for disks with masses in the range $0.075-0.1 M_{\odot}$.
For the same value of $Q_{min}$, lighter and colder disks produce clumps 
with appreciably smaller mass; the minimum scale of fragmentation is 
indeed set by the local Jeans mass, and it can be shown that this 
scales as $T^{5/4}$ for a fixed value of $Q_{min}$ 
(see MA03). The higher the mass resolution (higher number
of particles) the higher is the number of gravitationally bound condensations
that arise. On the other end, $Q_{min} \sim 1.4$ marks the threshold between
fragmenting and non-fragmenting disks in a way that is independent on
resolution; disks with $Q_{min}=1.5$ or higher were evolved with 
increasing number of particles, always formed strong spiral patterns
but these never broke up into clumps. The degree of fragmentation depends 
very weakly on the magnitude of the coefficients of artificial 
viscosity; there is a trend of stronger fragmentation with lower
viscosity but once again this does not affect the threshold $Q_{min}$ 
(see MA02).

We investigate how the outcome of our simulations depends on the way we
set up the initial conditions by running a test in which an initially 
very light disk ($M=0.0085 M_{\odot}$) is grown in mass at a constant
rate over tens of orbital times until it reaches the same (ten times
bigger) mass of one of our standard disk models undergoing clump formation
(the temperature is kept fixed and is equivalent to that used for 
the latter model).
Fragmentation occurs even in such growing disk once the outer regions
approach $Q_{min}=1.4$ (Figure 1); this shows that weaker non-axisymmetric 
torques occurring at higher values of $Q$ while the disk grows do not
lead to self-regulation of
the disk at values of $Q$ higher than 1.4 through mass
redistribution. The results of this experiment weaken considerably one 
of the arguments against gravitational instability, namely that in real 
disks spiral instabilities would always saturate before fragmentation 
becomes possible (Laughlin \& Rozyczka, 1996).
A few simulations with 200,000 particles were carried out for as many as 30
orbital times to probe the evolution of the system of protoplanets up
to about 1000 years. Several mergers occur over a few orbital times after
the fragmentation has started, leaving 2 or 3 protoplanets with masses between
$0.7$ and $6 M_J$ on orbits with eccentricities in the range $e=0.1-0.3$.
The surviving protoplanets continue to accrete mass at a rate which is strongly
dependent on the equation of state adopted in this later part of the evolution.
For adiabatic conditions the accretion is negligible and the values we
just quoted should well represent the final masses of the planets, whereas
for isothermal conditions
the accretion rate can be as high as $10^{-5} M_{\odot}$/yr, 
and so protoplanets
can reach brown dwarf-like masses in a few thousand years (the latter is the
estimated lifetime of the disk before it is accreted onto the central star
due to the strong gravitational torques, see MA03).

\begin{figure}
\epsscale{1}
\plotone{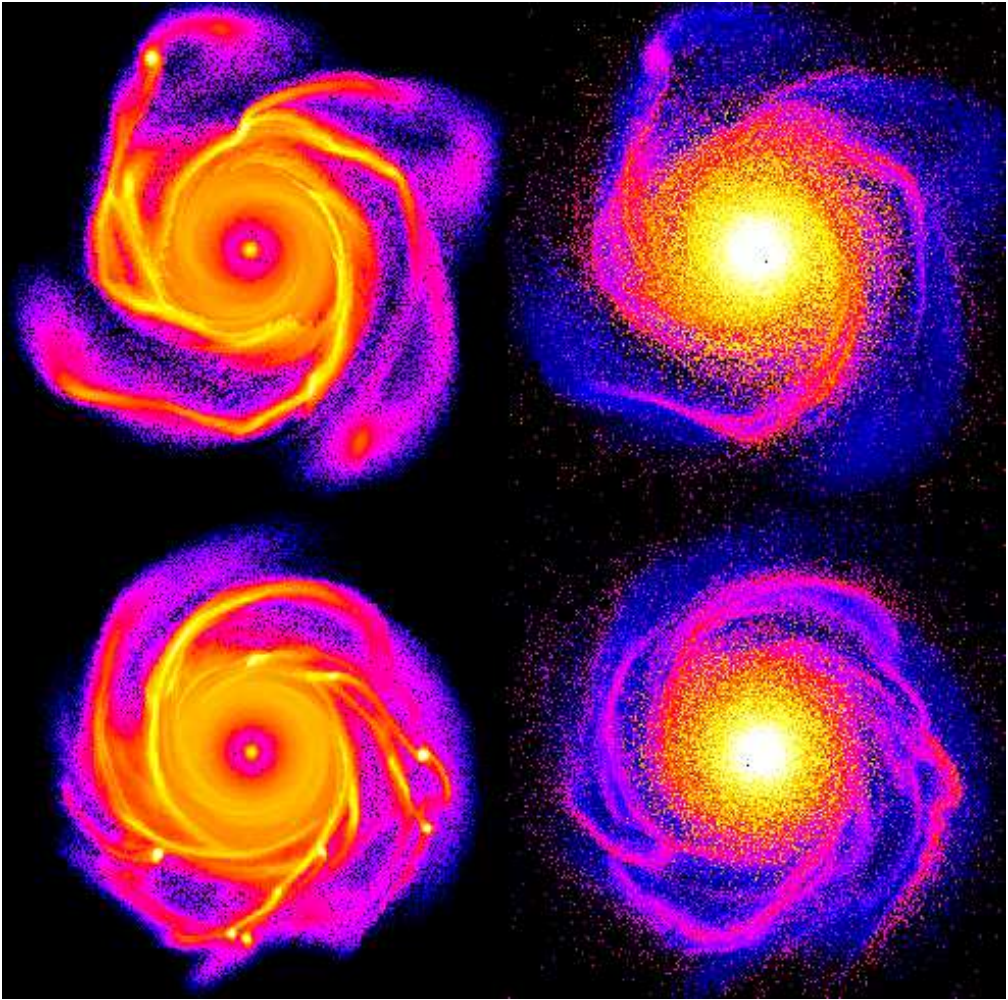}
\caption{Disk simulations with cooling and heating. Logarithmic density 
(left) and linear temperature maps (right) are shown for two simulations 
employing $10^6$ gas particles in the disk, after 300 years of evolution,
On top, the disk has a mass $M=0.1 M_{\odot}$, and we used $\gamma=5/3$ and 
$A=5$ (see section 2 and 3). At the bottom, the disk has a mass $M=0.085 
M_{\odot}$, and we used $\gamma=7/5$ and $A=3$.
Brighter colors correspond to higher values of the variables plotted.
Note the heating due to shocks along the spiral arms and the higher 
temperatures at the location of the protoplanets (up to 300 K).Boxes
are 25 AU on a side.}
\label {fig2}
\end{figure}

We also performed a number of runs in which disks are evolved with an adiabatic
equation of state since the very beginning. We explore different initial
$Q_{min}$ and different values of $\gamma$, in the range $1.2-1.4$. 
We find fragmentation only in runs starting from a very massive
disk ($M=0.125 M_{\odot})$ with unrealistically low
values of $Q$, $Q < 0.9$, and only for $\gamma = 1.3$ or lower. 
In particular only
if $\gamma=1.2$ do clumps survive for several orbital times and become
gravitationally bound. At these low values of $Q$ mass redistribution and 
shock heating are indeed so efficient that the disk quickly departs from
the initial conditions; such configurations are therefore unrealistic.

\section {Runs with radiative cooling and heating}.

In these runs ($10^6$ particles)
fragmentation is obtained for sufficiently short cooling timescales
(Figure 2). For any
given initial disk model we first run a simulation with a long cooling 
timescale which is expected to yield a stable configuration (see Rice
et al. 2003a) and then we re-simulate the same disk with increasingly 
smaller cooling timescales until we enter a regime where fragmentation
takes place. We find that
the latter occurs for critical cooling
timescales in the range $0.3-1.5 T_{orb}$ 
depending on disk mass (this varies between  $0.085$ and $0.1 M_{\odot}$)
and on the value of $\gamma$ used in the equation of state. 
At larger masses the higher disk self-gravity can amplify non-axisymmetric
perturbations more effectively and more rapidly, so lower cooling rates
are needed to counteract heating from strong compressions and shocks 
along the spiral arms.
For values of $\gamma=5/3$ our results are in very nice agreement with those
of Rice et al. (2003a,b), who used lower resolution SPH simulations and
a different setup of the initial conditions; we both find that $T_{cool}
\sim 0.8 T_{orb}$ or smalller is necessary to trigger fragmentation in
a disk with a mass $M=0.1 M_{\odot}$ (see Figure 2).
Instead, the critical cooling timescale for the same disk 
rises by more than  $50\%$  if $\gamma=7/5$ like in Boss (2002),
supporting his claim that convective transport of energy
at a rate comparable to the orbital time is enough to sustain the
instability.
Like in the runs described in the previous section, overdense regions
grow rapidly in mass and reach fragmentation provided that $Q_{min}$ 
drops below $1.4$. 
When comparing the same disk models, while most features like the
type of spiral pattern and the number of clumps formed are comparable
in these radiative runs and in those
started with a locally isothermal equation
of state, the spiral arms appears thinner and more filamemtary
in these new runs, owing to sharper density and pressure gradients.
The different density profile across the spiral arms is 
probably due to the fact that shock heating begins
as soon as the first non-axisymmetric structure appears in the new
runs, whereas it is completely inhibited below the critical density 
threshold in the locally isothermal runs.

We note that
disks that undergo fragmentation in the simulations of Rice et al. (2003a,b)
do that in a much stronger fashion compared to ours,
with several tens of gravitationally bound clumps instead of 
the few (between 1 and 10, see Figure 2) that survive the first 
violent phase of the 
gravitational instability  in our runs. This difference certainly arises
because we shut off cooling in the overdense regions once they grow
beyond the density threshold, while Rice et al. do not. In our runs the 
sites of formation of the protoplanets
coincide with those of the strongest pressure and temperature 
gradients along the spiral arms (see Figure 2), thus it is
not surprising that only the highest overdensity
peaks survive (see also Pickett et al. 2003); 
later, while the instability starts to fade away, these
few gravitationally bound clumps will be able to survive for long
timescales quite irrespective of the thermodynamical scheme adopted
because the disk enters a more quiet phase of its evolution 
and strong compressions do not occur anymore (see MA03).
However, even during the subsequent evolution we expect that
allowing or not allowing cooling within the densest regions will make
a difference; clumps
would contract nearly isothermally if strong cooling is 
always active and will reach higher mass
concentrations and thereby smaller effective sizes compared to the case 
in which cooling has been inhibited (see also Figure 2 of MA03 that
compares two runs with a locally isothermal equation of state, one with and
one without the switch to adiabatic later in the evolution).
A different size implies a different cross section for mergers, and
therefore a different mass spectrum and number of the clumps
after many orbital times.

Mass accretion rates of protoplanets are quite close to those found in previous
runs in which the disk was evolved adiabatically  
after fragmentation (see MA03).
One simulation with $10^6$ particles was carried out
for 600 years (equivalent to more than 20 orbital times at 10 AU); in that we
measured accretion rates $< 10^{-6} M_{\odot}/$yr during the late stage,
although there is considerable scatter when we look at the individual 
``histories'' of the protoplanets --- protoplanets that venture inside
10 AU are heated considerably and they can even lose some mass at the 
pericenter of their orbit, whereas those that spend most of the time at 
$R > 10$ AU have the highest accretion rate since they easily sweep the 
cold gas along their trail.

\section{Conclusions}

Gravitational instability continues to remain a very attractive mechanism
to explain the origin of gas giants, especially those found in extrasolar
planetary systems. With our new runs we showed that the mass range of
bound condensations arising along the spiral arms is significantly broad
once the relevant disk parameters are changed; the outcome is
not necessarily ``SuperJupiters'', protoplanets even as small as Saturn can be
formed. The general picture is confirmed with runs that directly model
radiative heating and cooling; the only necessary requirement for the
instability to proceed is that the cooling time must be comparable or 
only slightly larger than the orbital time at some point of  disk evolution.
Once protoplanets are formed, mergers during the first few orbital times 
and accretion of disk gas over longer timescales are the two ways by which 
they can grow in mass. How much they can grow will depend a lot on 
the details of heating and cooling. Our results
suggest that, even if the cooling time remains comparable to the orbital
time in the outer disk, a mass of $\sim 10 M_J$ might be an upper limit if
gravitational torques cause the dissipation of most of the disk on timescales 
of about $10^4$ years (see also MA03).

\end{document}